\documentclass[twocolumn,prb,preprintnumbers,superscriptaddress,showpacs]{revtex4}

\usepackage{graphicx} 
\usepackage{amsmath,amsfonts,amssymb,amsxtra}
\usepackage{siunitx} 
\usepackage{bm} 
\usepackage{upgreek} 
\usepackage{booktabs} 
\usepackage{color}


\newcommand{\e}{\mathrm{e}}
\newcommand{\ii}{\mathrm{i}}
\newcommand{\mwo}{MnWO$_4$}

\begin{document}

\textheight 24.40 true cm

\title{Kinetics of the Multiferroic Switching in MnWO$_4$}

\author{M.~Baum}
\affiliation{II.~Physikalisches Institut, Universit\"at zu K\"oln, Z\"ulpicher Str. 77, 50937 K\"oln, Germany}
\author{J.~Leist}
\affiliation{Institut f\"ur Physikalische Chemie, Georg-August-Universit\"at G\"ottingen, Tammannstr. 6, 37077 G\"ottingen, Germany}
\author{Th.~Finger}
\affiliation{II.~Physikalisches Institut, Universit\"at zu K\"oln, Z\"ulpicher Str. 77, 50937 K\"oln, Germany}
\author{K.~Schmalzl}
\affiliation{Juelich Centre for Neutron Science JCNS,
Forschungszentrum Juelich GmbH, Outstation at ILL, 38042 Grenoble,
France }
\author{A. Hiess}
\thanks{now at: ESS AB, Lund, Sweden}
\affiliation{Institut Laue-Langevin, BP 156, 38042 Grenoble Cedex
9, France}
\author{L.\,P.~Regnault}
\affiliation{Institut Nanosciences et Cryog\'enie, CEA-Grenoble, 38054 Grenoble Cedex 9, France}
\author{P.~Becker}
\author{L.~Bohat\'y}
\affiliation{Institut f\"ur Kristallographie, Universit\"at zu
K\"oln, Greinstr. 6, 50939 K\"oln, Germany}
\author{G.~Eckold}
\affiliation{Institut f\"ur Physikalische Chemie, Georg-August-Universit\"at G\"ottingen, Tammannstr. 6, 37077 G\"ottingen, Germany}
\author{M.~Braden}%
\email{braden@ph2.uni-koeln.de}%
\affiliation{II.~Physikalisches Institut, Universit\"at zu K\"oln, Z\"ulpicher Str. 77, 50937 K\"oln, Germany}

\date{\today}
\begin{abstract}

The time dependence of switching multiferroic domains in MnWO$_4$
has been studied by time-resolved polarized neutron diffraction.
Inverting an external electric field inverts the chiral magnetic
component within rise times ranging between a few and some tens of
milliseconds in perfect agreement with macroscopic techniques.
There is no evidence for any faster process in the inversion of
the chiral magnetic structure. The time dependence is well
described by a temperature-dependent rise time suggesting a
well-defined process of domain reversion. As expected, the rise
times decrease when heating towards the upper boundary of the
ferroelectric phase. However, switching also becomes faster upon
cooling towards the lower boundary, which is associated with a
first-order phase transition.

\end{abstract}

\maketitle

\section{Introduction}

Magnetoelectric materials allow one to tune both the electric
polarization by an external magnetic field as well as the magnetic
polarization by an electric field \cite{1,2}. In particular, the
control of magnetic order by an electric field has large
application potential in the context of data storage, but in spite
of strong efforts no suitable materials have been discovered for a
long time \cite{2,3}. In the recently discovered multiferroic
transition-metal oxides ferroelectric polarization arises from a
complex -- in most cases chiral -- magnetic structure. The
ferroelectric polarization can be modified by an external magnetic
field \cite{3,4,4b,4c,4d} in these multiferroic materials.  The
opposite direction is more difficult to study,  as the complex
antiferromagnetic order requires a microscopic technique directly
probing the spin arrangement \cite{6, 6b, 6c, 6d, 6e}. Using
polarized neutron scattering it has been shown in TbMnO$_3$
\cite{5}, LiCu$_2$O$_2$ \cite{5b} and in MnWO$_4$ \cite{5c} that
the chiral component (i.e.\ sense of rotation) of the magnetic
order  can be poled by an electric field when cooling through the
ferroelectric transition. However, the direct observation of the
electric-field induced switching of the chiral magnetism in these
spiral multiferroics has only been recently observed in
measurements on multiferroic MnWO$_4$ \cite{8a, 8} and
Ni$_3$V$_2$O$_8$ \cite{8b}, where full hysteresis cycles were
recorded.

The space group of MnWO$_4$ is $P2/c$ ($a =\SI{4.823}{\angstrom}$,
$b = \SI{5.753}{\angstrom}$, $c= \SI{4.992}{\angstrom}$, $\beta =
\SI{91.08}{\degree}$ at 300\ K). MnWO$_4$ undergoes a sequence of
magnetic phase transitions \cite{7}. Below 13.5\,K an
incommensurate spin density wave with collinear moments in the
$ac$-plane and propagation vector $\bm k = (-0.241, \frac{1}{2},
0.457)$ sets in (AF3). Below 12.3\,K an additional $b$-component
evolves and the moments order in an elliptical spiral (AF2). Below
~7.5\,K \cite{comment} the system orders again collinearly with
moments in the $ac$-plane but with a commensurate propagation
vector $\bm k = (-\frac{1}{4}, \frac{1}{2}, \frac{1}{2})$ (AF1)
\cite{7}.  At the transition to the non-collinear state (AF3
$\rightarrow$ AF2) spontaneous electric polarization parallel $b$
develops continuously. At the transition back into the collinear
but anharmonic state (AF2 $\rightarrow$ AF1) the electric
polarization disappears discontinuously \cite{9, 10, 11}.

\begin{figure}[t]
\includegraphics[width=0.75\columnwidth,angle=0]{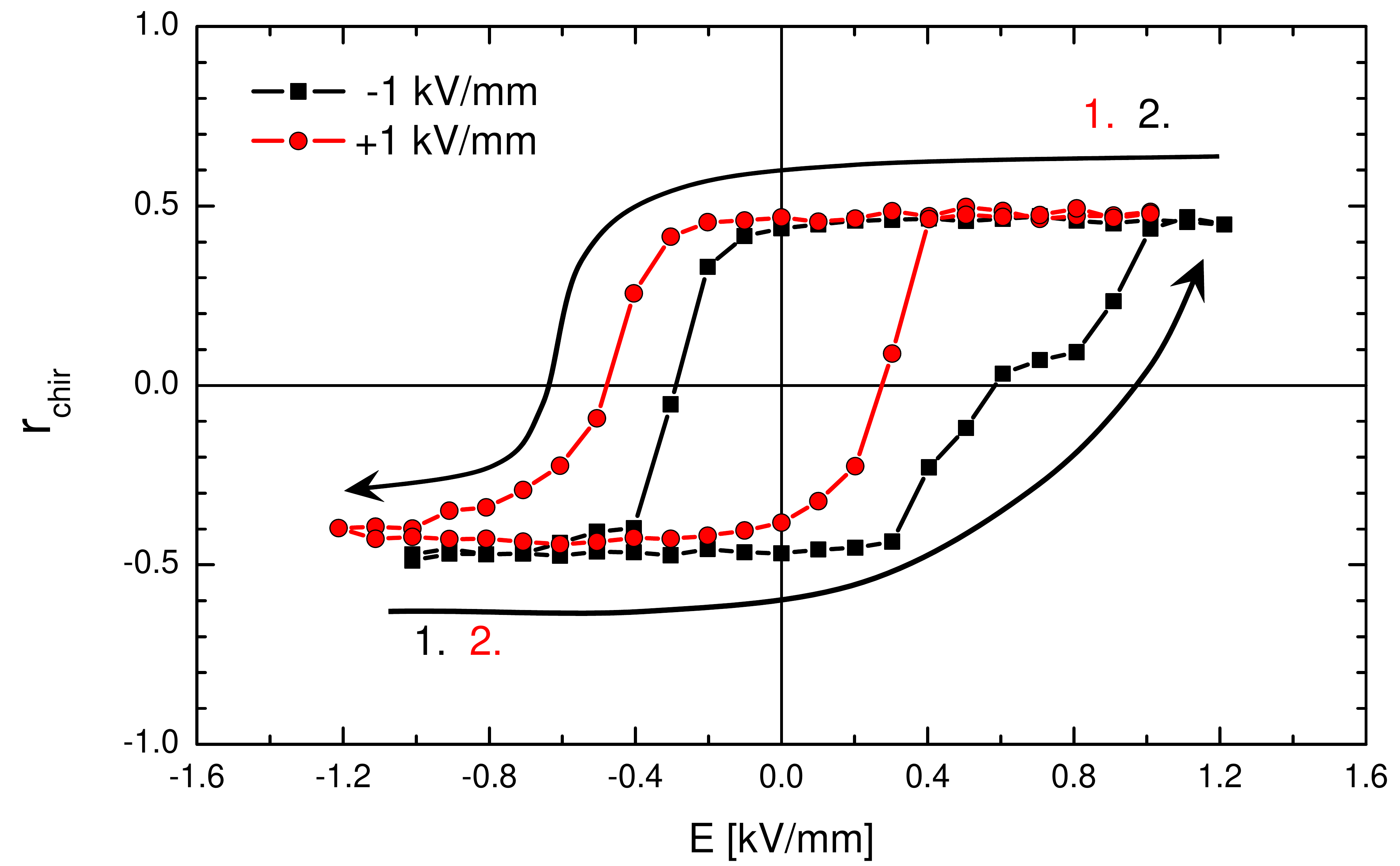}
\caption{Static hysteresis loop obtained by measuring the chiral
ratio defined by equation (2) as a function of external electric
field at constant temperature for $\bm Q = (-0.241, \frac{1}{2},
0.457)$. The loops were recorded after field-cooling from 20\,K to
10\,K in $ E = \SI{\pm1}{kV/mm}$. Note that the coercive field
depends on the field applied during cooling. Arrows indicate the
sequence of the hysteresis in the two cycles after cooling the
sample in positive and negative electric field, respectively.}
\label{hys}
\end{figure}

Multiferroicity in MnWO$_4$ can be explained by the inverse
Dzyaloshinskii-Moriya interaction \cite{katsura}. The direction of
the electric polarization $\bm P$ is given by $\bm P \propto \bm
e_{ij} \times (\bm S_i \times \bm S_j)$ where $\bm S_i$ and $\bm
S_j$ are the magnetic moments of  manganese ions and $\bm e_{ij}$
points along the connection line of the corresponding ions
\cite{9}. Summing up these contributions over all pairs of
magnetic ions yields the total ferroelectric polarization, which
in case of \mwo \ points along the $b$ direction.

While the microscopic coupling leading to multiferroic order seems
well understood in MnWO$_4$ as well as in the other chiral
multiferroics, very little is known about the dynamics of the
domain inversion. In this work we present investigations on the
kinetics of electric-field induced switching of chiral magnetic
structures in multiferroic MnWO$_4$. We applied a stroboscopic
technique combined with polarized neutron diffraction in order to
investigate how fast the magnetic chirality adapts to an
instantaneously switched electric field.

\section{Experimental}

\subsection{ General aspects of neutron scattering on chiral
multiferroics}

Scattering experiments with polarized neutrons are ideally suited
for studying magnetic structures with a non-vanishing term $\bm
S_i \times \bm S_j$, as they give access to the so-called chiral
term: $-\ii(\bm M_\perp \times \bm M_\perp^*)_x$. We use the
common right-handed coordinate system with orthogonal axes: $-\bm
x \, \| \, \bm Q = \bm k_\mathrm{i} - \bm k_\mathrm{f}$, $\bm z$
vertical, $\bm y = \bm z \times \bm x$. $\bm M_\perp(\bm Q)$ is
the part of the three-dimensional magnetic structure factor (i.e.\
the Fourier coefficient of the magnetization density) $\bm M(\bm
Q)$ perpendicular to the scattering vector $\bm Q$. We define the
chiral ratio as the quotient of chiral and total magnetic
scattering at a given $\bm Q$:

\begin{equation}
 r_{\mathrm{chir}} = \frac{-\ii(\bm M_\perp
\times \bm M_\perp^*)_x}{|\bm M_\perp|^2}. \end{equation}

The chiral ratio can amount to $\pm$1 in the case of an ideal
helix and scattering vector parallel to the propagation vector,
but in general it will be smaller even for the ideal helix. The
chiral ratio for a mono-domain sample can easily be calculated
from the magnetic structure \cite{8}. The deviation with the
experiment then directly gives the distribution of the chiral
domains.

With spherical polarization analysis it is possible to measure the
scattering intensity $I_{ij}$ for any directions of the incoming,
index $i$, and outgoing, index $j$, neutron polarization.
Over-bars indicate antiparallel neutron polarization. Therefore
also the rotation of neutron polarization can be analyzed. The
chiral ratio can be measured equally well in several channels of
the neutron polarization matrix \cite{brown,8}. It can be detected
in the spin-flip scattering with polarization parallel to the
scattering vector: $I_{x\bar{x}} = |\bm M_\perp|^2 - \ii(\bm
M_\perp \times \bm M_\perp^*)_x$ and $I_{\bar{x}x} = |\bm
M_\perp|^2 + \ii(\bm M_\perp \times \bm M_\perp^*)_x$ yielding

\begin{equation}
r_{\mathrm{chir}} = \frac{I_{x\bar{x}} -
I_{\bar{x}x}}{I_{x\bar{x}} + I_{\bar{x}x}},
\end{equation}

 as well as in transverse polarization channels \cite{8}.

\begin{figure}[t]
\includegraphics[width=0.999\columnwidth, angle=0]{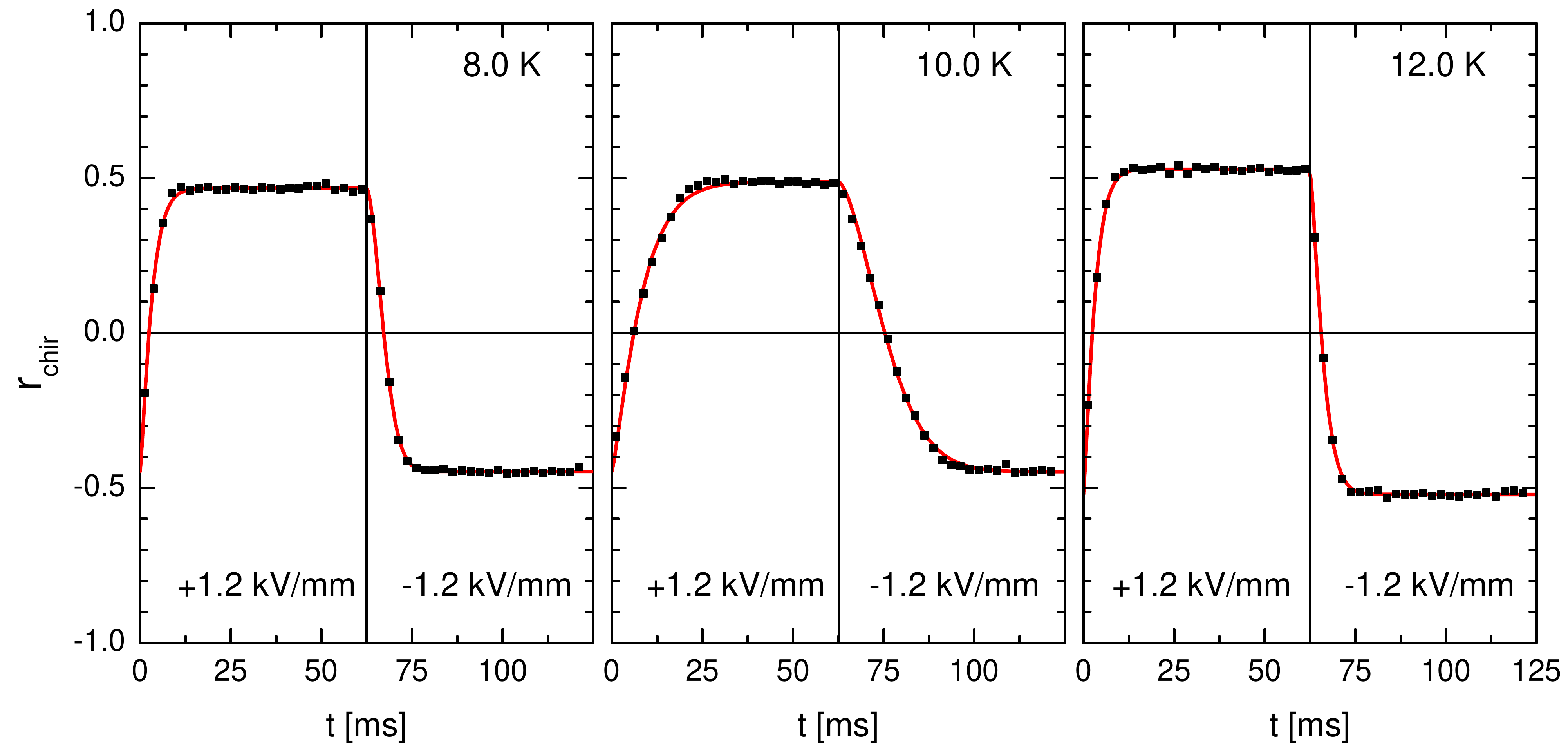}
\caption{Time dependent measurement of the magnetoelectric
switching at different temperatures for $\bm Q = (-0.241,
\frac{1}{2}, 0.457)$. The electric field is switched with a
frequency of 8\,Hz and an amplitude of $\SI{\pm1.2}{kV/mm}$, rise
time 0.2\,ms. The chirality can approximately be switched in the
same range as in the quasistatic hysteresis loop.  The rise times
differ for the two states. Before recording the time dependencies
the sample was cooled from 20\,K in an electric field of
+1.2\,kV/mm.} \label{schalten}

\bigskip

\includegraphics[width=0.8\columnwidth, angle=0]{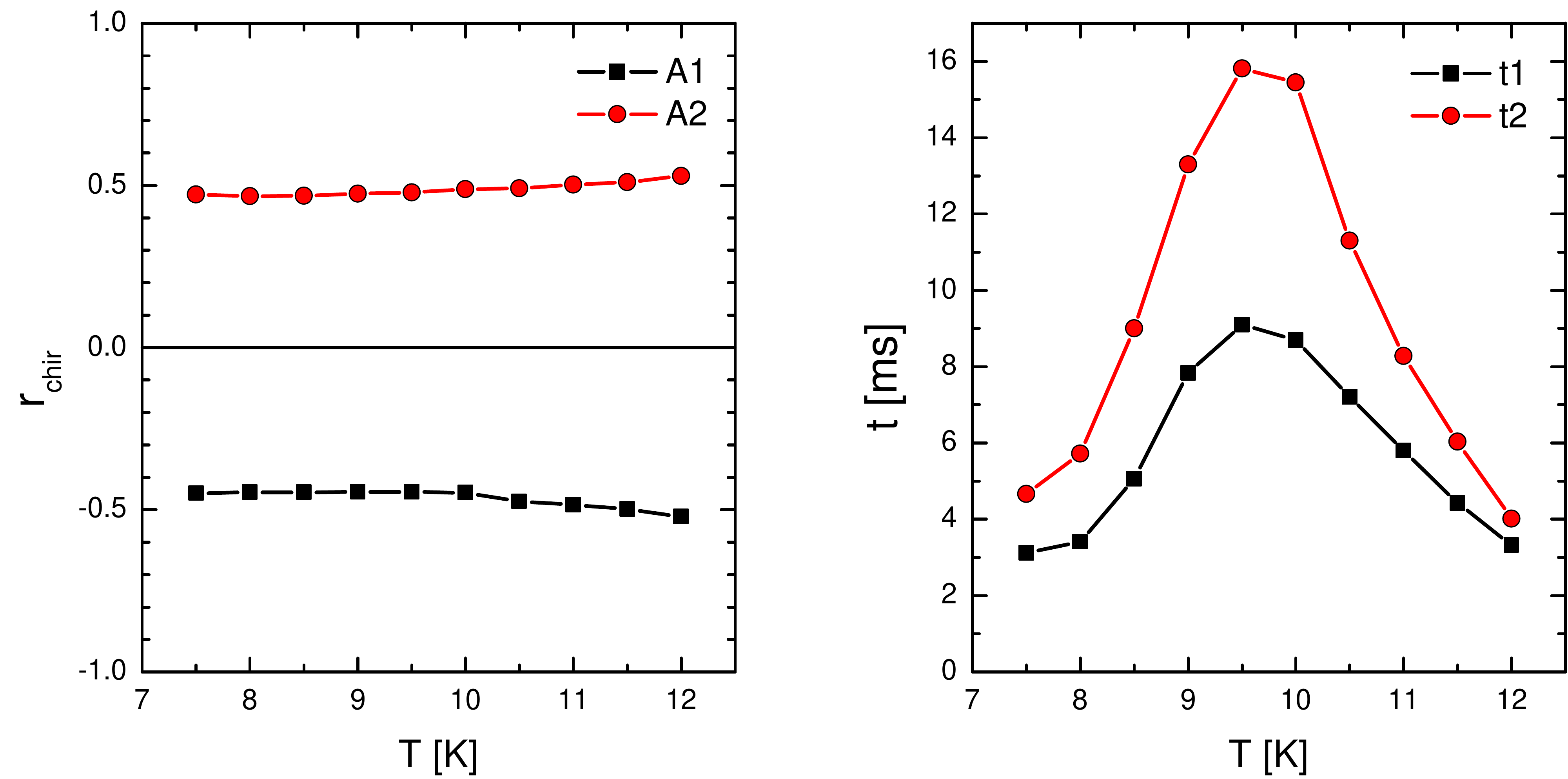}
\caption{Characteristics of the switching behavior for field
cooling the sample with $E=+1.2$\,kV/mm. The left and right panels
give the amplitudes and the relaxation times, respectively, that
were obtained by fitting the stretched exponential function,
equation (3), to the stroboscopic data. 1 and 2 denote the
relaxation in positive and negative external fields (first and
second half of the time profile), respectively. The electric field
is switched with a frequency of 8\,Hz and an amplitude of
$\SI{\pm1.2}{kV/mm}$. The exponents $b_1$ and $b_2$ were fitted
globally for the whole temperature range: $b_1 = 1.21(1)$, $b_2 =
1.63(2)$. The temperature was decreased during the measurement. }
\label{werteschalten}
\end{figure}

\subsection{ Polarized neutron scattering experiments on MnWO$_4$}

Neutron scattering experiments were performed on the two cold
triple-axis spectrometers IN12 and IN14 at the Institut Laue
Langevin (ILL). On both instruments a bender was set between the
monochromator (pyrolitic graphite) and the sample in order to
polarize the incoming neutron beam. The polarization analysis of
the scattered neutrons was performed with a Heusler crystal. For
the control of the neutron spin at the sample position a
\textsl{Cryopad\,III} was installed on the IN12 spectrometer. In
total a high degree of polarization was achieved. The flipping
ratios measured on nuclear Bragg reflections (0 2 0) and (1 0 -2)
amount to 26 and 41, respectively. The slightly lower precision of
the polarization at lower scattering angle most likely arises from
perturbations in the guide fields. For the second experiment on
IN14 a Helmholtz-coil setup was used to control the neutron
polarization which only allows longitudinal polarization analysis.
For analyzing the chiral domains longitudinal polarization is
sufficient, as the chiral contribution can be isolated in the $xx$
channel.

\subsection{Stroboscopic technique for neutron scattering}

The idea of our measurement was to reverse an electric field  and
to detect the response of the magnetic structure as a function of
time. Typical count rates in neutron scattering experiments range
in the time-scale of seconds (elastic) to minutes (inelastic
scattering). Therefore studying the time dependence of a single
process would not yield enough counting statistics. To overcome
this challenge stroboscopic neutron-scattering techniques   were
developed \cite{12}. With this method it is possible to
synchronize a periodical perturbation at the sample with the count
rate of the neutron detector. The signal from the detector is
recorded in time slots by a multichannel scaler. In order to gain
sufficient count rates for each time slot the measurement is
repeated periodically while the count rates of the individual time
slots are accumulated. This technique was installed on the cold
triple-axis spectrometer \textsl{IN12} at the \textsl{Institut
Laue-Langevin} using the \textsl{Cryopad\,III} setup for spherical
polarization analysis. For the study of the multiferroic domain
switching in \mwo \ we applied a periodic electric field with
rectangular time shape of amplitudes of about $\SI{\pm1.5}{kV/mm}$
and half periods of typically 67.5\ ms. The electric field was
applied by setting the flat single crystals (thickness of the
order of 1-2\ mm) between two aluminium plates, which were
connected with the switchable high-voltage generator. The time
slots were in the range of ms. With the fast electronics and with
the signal strength of about 1000 cnts/s, it is possible to get
sufficient statistics for the entire time dependence within
several minutes to an hour.

\section{Kinetics of multiferroic domains studied by neutron
scattering}

\subsection{Temperature dependence of the multiferroic rise times }

\subsubsection{First set of experiments on the IN12 spectrometer }

In the first experiment the domain kinetics was analyzed on the
IN12 spectrometer. By putting the sample between two large
aluminium plates the electric field was applied along the
crystallographic $b$-axis, which is the direction of the
spontaneous electric polarization \cite{7,9,10}. The sample
thickness amounts to 1.98\,mm in this direction. The sample was
mounted in the $(0,1,0)/(-0.214, 0, 0.457)$ scattering-plane in
order to reach the incommensurate magnetic Bragg peaks.  First we
recorded static hysteresis loops in order to determine the field
strength needed to reverse chiral domains and the chirality
saturation that could be achieved, see Fig.~\ref{hys}. At the
magnetic Bragg peak $(-0.241, \frac{1}{2}, 0.457)$ the hysteresis
loops were recorded after cooling the sample from 20\,K to 10\,K
in an applied electric field of $\pm1$\,kV/mm. The maximal
chirality is $\pm0.48$ and can be fully reversed. We find that the
crystal develops a preferred chirality depending on the field
direction which is applied during cooling from the paramagnetic
phase. A higher field is needed to force the sample in the
not-preferred state and a lower field can reverse it into its
preferred state. This is in accordance with previous results on
the control of multiferroic domains by varying an electric field
at constant temperature\cite{8,13}.

We started studying the time dependence of the multiferroic domain
inversion by switching an electric field of $\SI{\pm1.2}{kV/mm}$
with a frequency of 8\,Hz and by registering the $I_{\bar{x}x}$
and $I_{x\bar{x}}$ intensities at the same magnetic Bragg peak as
for the hysteresis loops. We observed the domain relaxation in the
temperature range from 12.5\,K to 7.5\,K. The sample was first
cooled from 20\,K to 12.5\,K in a field of +1.2\,kV/mm. Three
time-resolved curves at different temperatures are shown in
Fig.~\ref{schalten}. We see that it is possible to switch the
magnetic chirality frequently between the two saturation values
and we reach the same amplitude as for the static hysteresis loop.
No evidence for a change in the domain response was seen in these
runs. Inspection of Fig.~\ref{schalten} immediately shows that the
rise time is shorter when the system goes into its preferred state
and longer when it is forced to the other state. At the
temperatures closer to the upper and the lower boundaries of the
ferroelectric phase the domains switch faster (i.\,e. the rise
times of the chiral ratios are smaller). This qualitative analysis
agrees with that obtained from other techniques studying the
reversal of the ferroelectric domains. Second harmonic generation
experiments also find a rather slow domain inversion at T=12\ K
\cite{hoffmann} as well as the comprehensive study by means of
dielectric spectroscopy \cite{niermann}. The dielectric response
and this neutron diffraction study show faster response near the
upper and the lower transitions. While this behavior is expected
for approaching the upper continuous transition, it is quite
unexpected for the lower first-order transition.

\begin{figure}[t]
\includegraphics[width=0.999\columnwidth, angle=0]{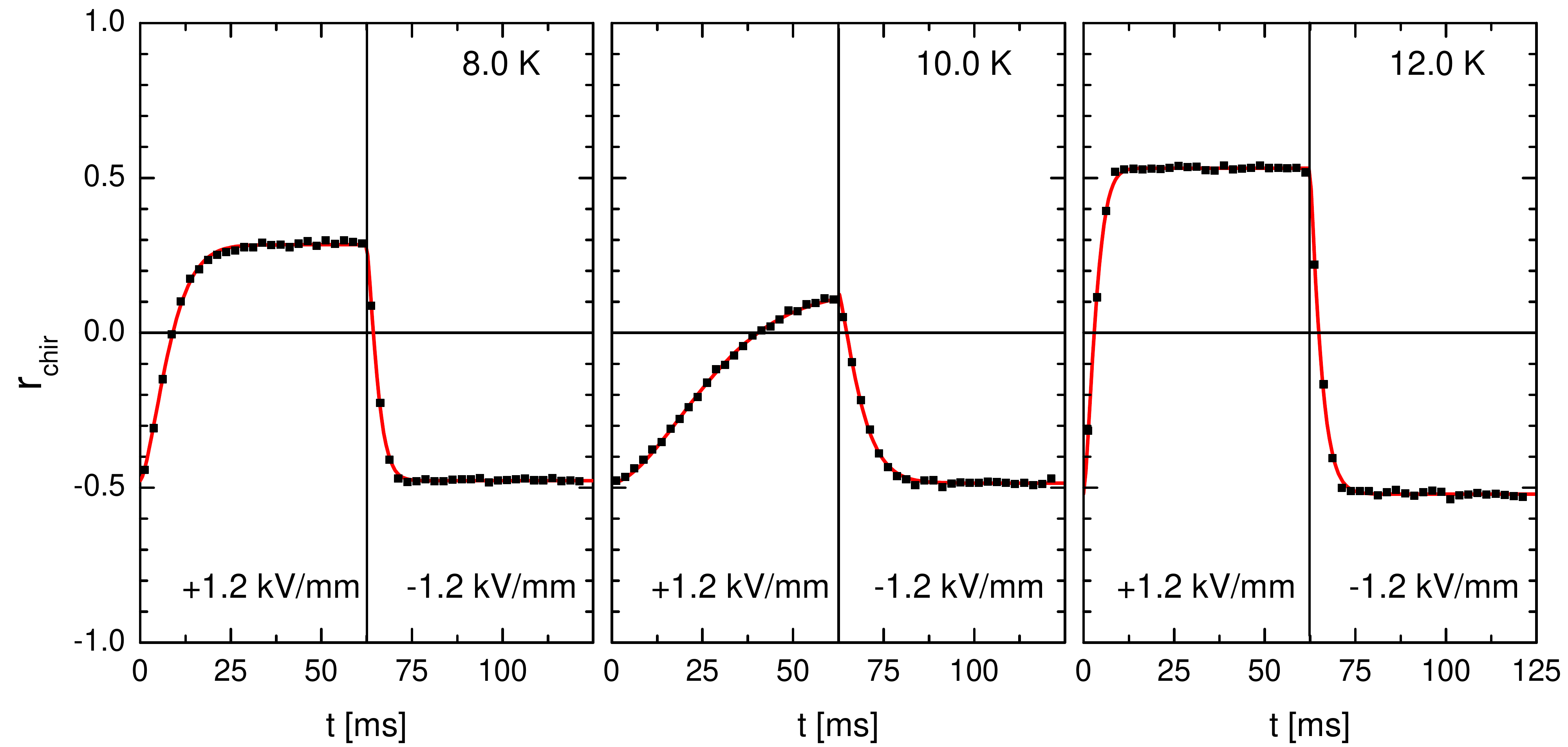}
\caption{Time dependent measurements of the multiferroic switching
at different temperatures. The sample was cooled from 20\,K in an
electric field of $\SI{-1.2}{kV/mm}$. Frequency: 8\,Hz, amplitude:
$\SI{\pm1.2}{kV/mm}$. The switching behavior is highly asymmetric.
} \label{umgepolt}

\medskip

\includegraphics[width=0.8\columnwidth, angle=0]{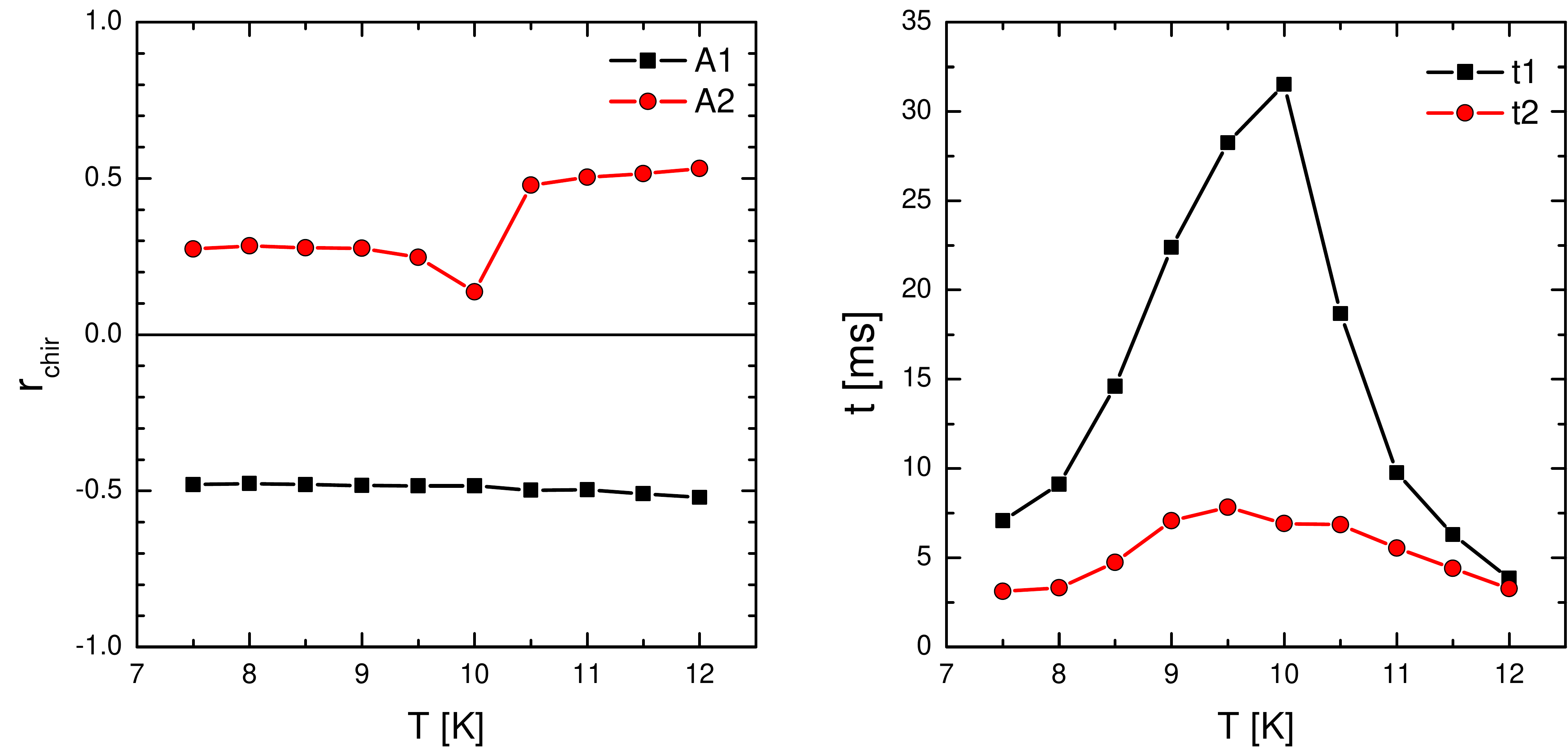}
\caption{Characteristics of the switching behavior for field
cooling $-1.2$\,kV/mm. Frequency: 8\,Hz, amplitude:
$\SI{\pm1.2}{kV/mm}$. The exponents $b_1$ and $b_2$ were fitted
globally for the whole temperature range: $b_1 = 1.63(1)$, $b_2 =
1.25(1)$. The temperature was decreased during the measurement. }
\label{werteumgepolt}
\end{figure}

In order to quantitatively analyze the dynamics of the domain
inversion, the following function was fitted to the data:

{\scriptsize
\begin{equation}
\begin{split}
y(t) =  &\frac{1}{2}\left[1-\tanh\left(\frac{t-t_0}{\SI{0.001}{ms}}\right)\right]\left[A_2+(A_1-A_2)\e^{-\left(\frac{t}{t_1}\right)^{b_1}}\right] \\ +&\frac{1}{2}\left[1+\tanh\left(\frac{t-t_0}{\SI{0.001}{ms}}\right)\right]\left[A_2+(A_1-A_2)\left(1-\e^{-\left|\frac{t-t_0}{t_2}\right|^{b_2}}\right)\right]
\end{split}
\label{relax}
\end{equation}}

\noindent The hyperbolic tangent yields a continuous approximation
of the heaviside step function which describes the instantaneous
reversal of the electric field. $t_0$ is half the period.
$t_{1,2}$ are the characteristic rise times for the two field
directions (positive and negative, respectively). $A_1$ and $A_2$
describe the minimal and maximal chiralities that are obtained for
the two field directions, respectively. We discuss the relaxation
on the basis of the Avrami model, which corresponds to a simple
domain growth \cite{14, 15}.
For the two directions of the field
inversion, the exponents $b_1$ and $b_2$ can be globally fitted
for the whole temperature range yielding values of 1.21 and 1.63
for the faster and for the slower switching, respectively. The
characteristic time constants and the maximal chiral ratios
recorded at  all temperatures are shown in
Fig.~\ref{werteschalten}.  The properties, which were already
discussed at the three example runs (Fig.~\ref{schalten}), are
confirmed by the whole temperature dependence.

As reported previously \cite{hoffmann,niermann}, the multiferroic
domain inversion in \mwo \ is rather slow, of the order of
milliseconds to a few tens of milliseconds indicating that the
domains in the multiferroic state are well pinned. This is in
accordance with the large coercive fields found in the
quasi-static experiments \cite{8,kundys}. The slow multiferroic
rise times underline the importance of the lattice in the control
of domains. For a purely structural domain the finite sound
velocity can limit the speed of domain inversion to below the
quotient of thickness to sound velocity (of the order of $\mu$s
for a mm sized crystal), but this is not the limiting factor in
\mwo . That we observe three orders of magnitude larger rise times
indicates effective pinning. Our neutron experiment senses the
chiral magnetic component that could in principle be decoupled
from the total ferroelectric polarization during the domain
inversion. The good agreement between the magnetic response and
the dielectric one \cite{niermann} clearly excludes such
decoupling. Furthermore we always find a single relaxation; there
is no indication for some faster magnetic relaxation preceding
slower ferroelectric domains. This documents once more, that
magnetism and ferroelectric polarization are strongly coupled in
\mwo .

The precise reason for the slow domain inversion remains unclear.
Since only tiny atomic displacements can be associated with the
observed small ferroelectric polarization, it seems unlikely that
strong pinning arises due to these purely ferroelectric
displacements. Instead higher-order modulations appear more likely
to be at the origin of the strong pinning. For a magnetic
modulation one can always expect a coupling with a structural
modulation of half the magnetic period. In reference \cite{13} it
was shown, that this half-period structural modulation is sizeable
in \mwo \ and that it is accompanied by a second-order modulation
of the magnetic structure. The two modulations exhibit coherent
interference underlining their tight coupling \cite{13}. This
structural distortion can possess a higher pinning potential than
the weak ferroelectric distortion itself. It is quite astonishing
that domains switch again much faster when approaching the lower
boundary of the multiferroic phase. Since this transition is of
clear first order, one would not expect any slowing down of
excitations, which could melt the pinning. Instead the lower
transition is related to the anharmonic distortions of the
magnetic modulation, which increase upon cooling in the
multiferroic phase \cite{13}. This increase is relatively stronger
than that of the first-order magnetic modulation. Increasing
anharmonicity may result in some depinning.

After this first series of experiments we tried to investigate
whether we could change the preferred chirality by reversing the
field during cooling. We heated the sample to 20\,K and cooled it
in a field of $\SI{-1.2}{kV/mm}$. Then we recorded the switching
behavior in a temperature range from 12.5\,K to 7.5\,K. Three
curves at different temperatures are shown in Fig.~\ref{umgepolt}.
As expected the sample now has a preference to negative chirality
what is in accordance with the field applied during cooling.
Comparing  the fit parameters of the two temperature runs
(Fig.~\ref{werteschalten} with Fig.~\ref{werteumgepolt}) reveals
reversed behavior. Now the negative chirality exceeds the positive
chirality. However the positive (not  preferred) chirality shows a
less smooth curve than before because the full saturation is not
reached at all temperatures. The behavior of the time constants is
also reversed. Now the time needed to switch to the positive
chirality exceeds the time needed to switch to the negative
chirality. We also observe that the rise time to switch to not
preferred chirality became longer than that observed in the first
series.

The reversal of the preferred orientation agrees with the
pronounced memory effects reported in reference \cite{8}. Most
likely the cooling with a large electric field freezes in some
local fields which determine the preferential orientation of the
multiferroic domains. This leads to the strong asymmetries in the
hysteresis cycles as well as in the time dependence of domain
reversion.

\subsubsection{Second set of experiments on the IN14 spectrometer using a different crystal }

\begin{figure}[t]
\includegraphics[width=0.85\columnwidth, angle=0]{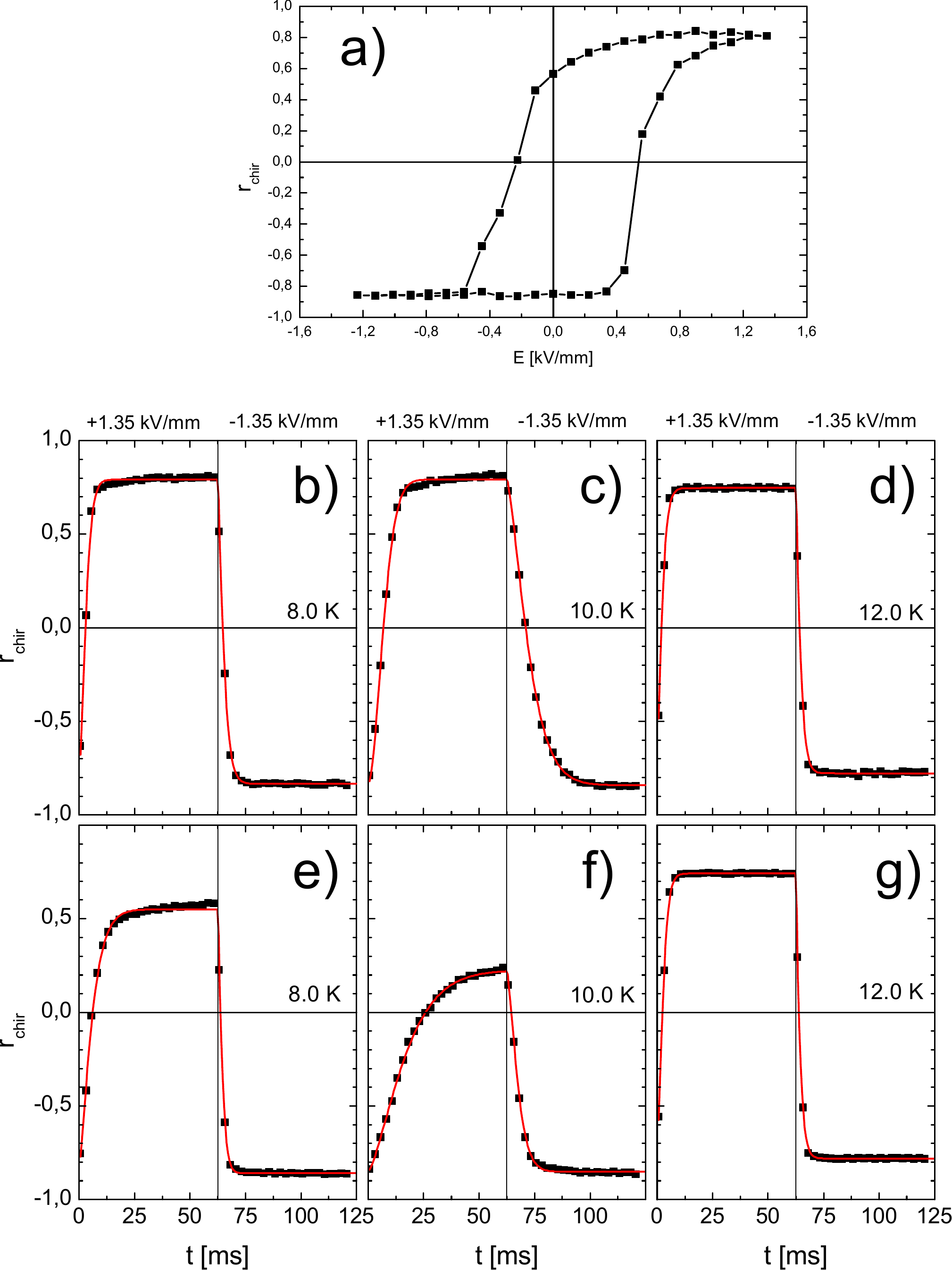}
\caption{ Experiments performed on the IN14 spectrometer with a
0.89 mm thick crystal of MnWO$_4$. The upper panel shows a
multiferroic hysteresis loop of the chiral ratio recorded after
zero-field cooling the crystal; note that the loop still exhibits
finite asymmetry. The middle and lower panels show the results of
time dependent measurements of the multiferroic switching. The
electric field is switched with a frequency of 8\ Hz and an
amplitude of $\pm$1.35\ kV/mm. The middle and lower rows show data
after field-cooling the sample from 20 K in an electric field of
+1.35\ kV/mm and -1.35\ kV/mm, respectively. } \label{in14data}
\end{figure}

The time-resolved measurement was repeated at the IN14
spectrometer at the ILL with a crystal grown in a different batch,
in order to investigate which properties were specific to the
sample and which ones are general. The second sample was 0.89 mm
thick. For this experiment we used a conventional Helmholtz setup
for the polarization analysis as it provides a higher scattering
intensity, and as the chiral component can be studied in the
longitudinal $xx$ channel. The intensity of the magnetic
reflection $(-0.214,0.5,0.457)$ amounts to 1400 cnts/s, i.e.\
comparable to the first experiment in spite of the smaller sample
volume. First, we recorded a hysteresis loop after zero-field
cooling the sample and thereby avoiding a freezing-in of electric
fields, see  Fig.~\ref{in14data} a). At 10 K a negative field of
$-1.25$ kV/mm was applied yielding a saturation value of the
chiral ratio of 0.85 which is significantly higher than that for
the first sample and indicates an almost perfect monodomain state
(95\% , comparable to earlier studies \cite{8}). Nevertheless, the
hysteresis loop is not perfectly symmetric but shows a clear
preference for negative electric polarization which corresponds to
the first field the crystal sensed after cooling. \mwo \ crystals
thus possess some intrinsic preference for the multiferroic state,
whose origin requires further clarification. For the
time-dependent studies we chose a field of $\pm$1.35 kV/mm, which
is sufficiently above the coercive fields in both directions.

The first set of time-dependent measurements was taken after
field-cooling the sample from 20 K to 12 K with an electric field
of +1.35 kV/mm. The electric field was modulated with a frequency
of 8 Hz and an amplitude of $\pm$1.35\ kV/mm, and time-resolved
data were recorded every 0.5\ K down to 7.5\ K. Three
characteristic time scans are shown in Fig.~\ref{in14data} b)-d).
The reproducibility of the results was verified by raising the
temperature to 20 K and field-cooling the sample again and by
taking data with 1\ K steps.  The obtained data are perfectly
reproducible in these two runs indicating in particular the
absence of fatigue, see Fig. 7.

\begin{figure}[t]
\includegraphics[width=0.9\columnwidth, angle=0]{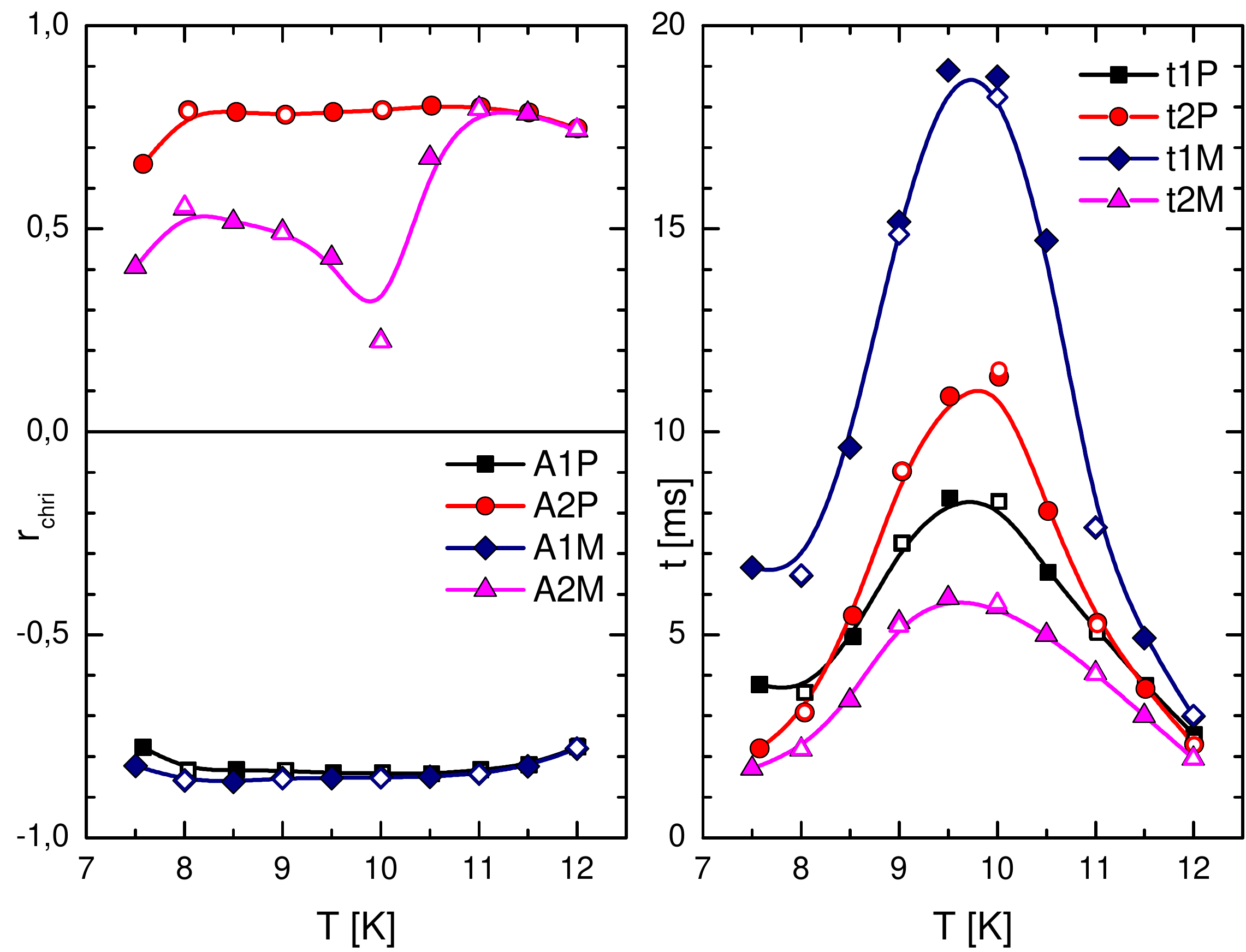}
\caption{Results of the analysis of the time dependent
measurements of the multiferroic switching in MnWO$_4$ performed
on the IN14 spectrometer with a thin sample. The left panel shows
the chiral amplitude of the hysteresis cycles, that could be
controlled as function of temperatures; the right panel shows the
rise times of these hysteresis cycles as function of temperature
for the two direction of electric-field inversion. Empty symbols
belong to the repeated measurement and proof the reproducibility
of the data. In total four temperature dependencies of the rise
time were performed corresponding to an initial field cooling in
positive or negative fields (labelled by $P$ and $M$ in the
legend) and to relaxation in positive and negative fields,
labelled by 1 and 2, respectively.} \label{in14fit}
\end{figure}

The entire procedure was repeated with this second crystal but
starting with field-cooling in a negative field of $-1.35$ kV/mm.
Characteristic time scans are shown in Fig.~\ref{in14data} e)-g)
and the results of the relaxation description are given in
Fig.~\ref{in14fit}. In this run the sample develops a clear
preference for negative fields and shows more asymmetric behavior.
The sign of the asymmetry agrees with the one observed after
zero-field cooling of the virgin crystal and therefore seems to be
intrinsic to this crystal. The rise times in the favorable
negative field direction of this run are faster than those in the
previous run, and the rise times in the unfavorable positive field
direction are even slower than the previous ones. So, the
field-cooling in negative fields seems to enhance the natural
asymmetry of this crystal, while field-cooling in the opposite
field seems to reduce it thereby yielding an almost symmetric
state. However, all the general aspects of the rise times with
maximum times near 10\ K are again observed. In this last run the
rise times in the unfavorable direction became too long to reach
saturation, therefore the amplitude in the unfavorable direction
diminishes towards low temperatures and exhibits a minimum around
10\ K.

The time-resolved data were again analyzed with the relaxation
formula (2). The exponents $b$ of the stretched exponentials could
be fitted globally for all temperatures or individually without
significant differences in the rise times; they amount to
1.67/1.32 and 1.43/1.36 in the slow/fast switching processes in
the first and second sequence, respectively, so that it actually
is a compressed exponential behavior. Although the stretching
exponent clearly depends on sample history and on the direction of
the field switch the average value agrees for the two samples and
the various histories studied. The average stretching exponent is
found at 1.4. This exponent can be interpreted within the
frame-work proposed by Ishibashi and Takagi \cite{15}. The
exponent corresponds to the dimensionality of the domain growth
enhanced by one for the case of constant continuous nucleation.
Therefore our data suggests a low-dimensional domain growth as it
is typically observed in ferroelectrics. A distribution of pinning
properties and of local rise times will decrease the fitted
exponent thereby weakening the conclusion of one-dimensional
domain growth, but a large exponent that is reduced by severe
spread of rise times is inconsistent with our time domain data.
%

In Fig.~\ref{in14fit} the rise times and amplitudes of all
measurements with the second setup are combined as function of
temperature. Qualitatively, these values agree with the results of
the first experiment, see Fig.~\ref{werteschalten} and
\ref{werteumgepolt}, but there is less difference between the up
and down field directions and the rise times are slightly shorter
in general. The controlled chiral component remains close to the
ideal value over the entire temperature range in the second
sample. Also this sample shows the faster domain response when
approaching the upper and the lower boundaries of the
ferroelectric phase. Additionally, the second sample displays a
weak long-time-scale relaxation process which can be observed at 8
K and 10 K in Fig.~\ref{in14data}. The main part of the chiral
ration adapts to the new field in less than 10 ms, but then the
chiral ratio still increases with a much lower time scale. This
behavior was not observed with the other sample studied on IN12,
which, however, could be switched to a lesser extent.

Due to the different initial poling (parallel and antiparallel to
the intrinsic preference of the crystal) and due to the two field
directions there are in total four different configurations to be
considered for each of the two crystals, see the combined plot in
the right panel of Fig. 7. If we average these four curves we
determine rise times of 2.5, 10.5 and 3.6\ ms at 12, 9.5 and 7.5\
K, respectively, for the second crystal, while the rise times of
the first crystal amount to 3.6, 15, and 4.5\ ms at the same
temperatures. The slightly slower domain reversion (roughly a
factor 1.5) in the first crystal can be attributed to the lower
applied field (see discussion below) so that the two sets of
measurements agree quite well with each other. By averaging the
four possible configurations one finds the intrinsic domain
properties for this type of crystals. Note, that the two crystals
were obtained from different growth processes but using the same
technique \cite{10}.


\subsection{Electric field and frequency dependence of the rise
times}

\begin{figure}[t]
\includegraphics[width=0.97\columnwidth, angle=0]{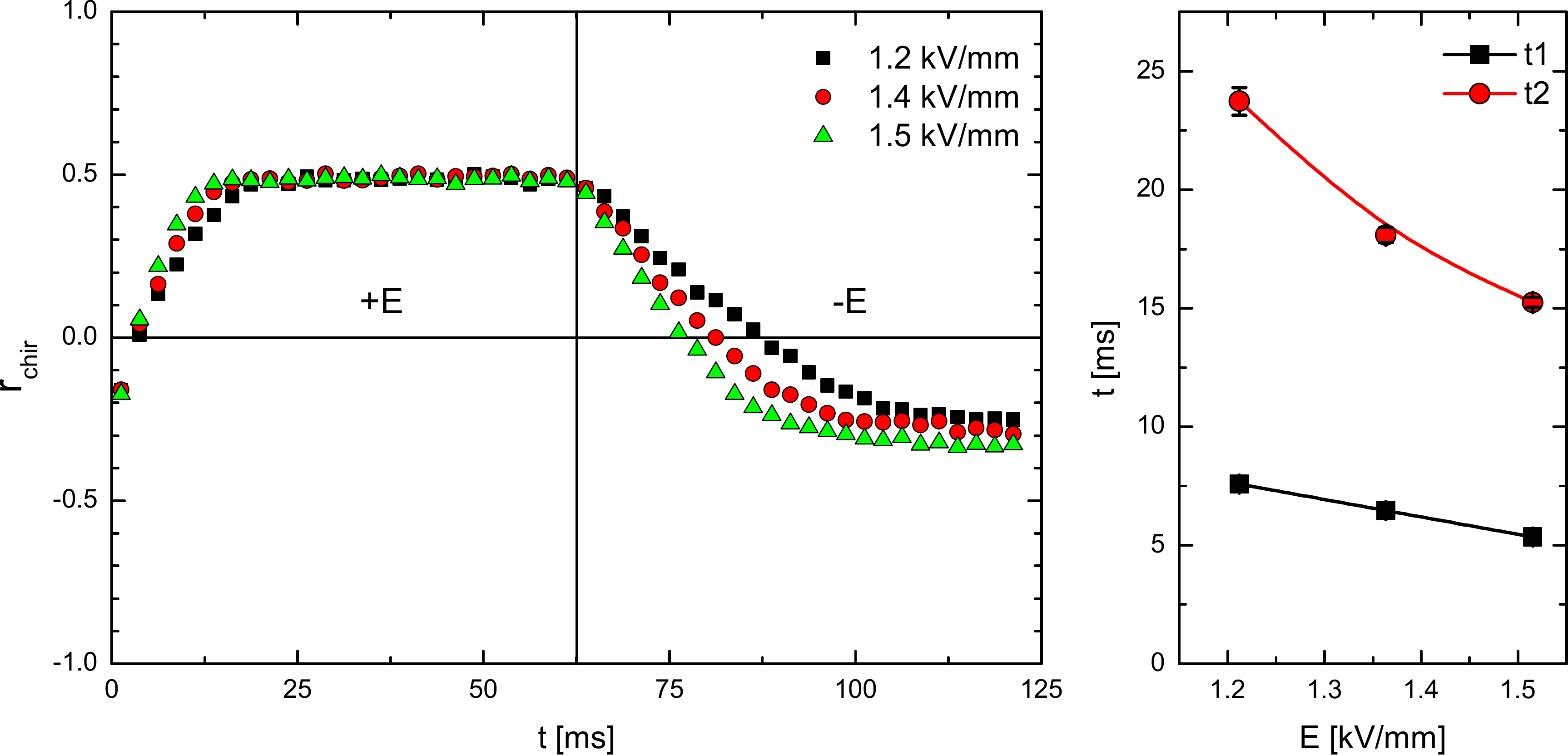}
\caption{Switching with different electric field strength at 10\,K
and 8\,Hz. The left panel shows the time-domain data obtained at
various fields. Higher fields yield a higher driving force and
thus a faster respond. The right panel gives the rise times as
function of external field (data taken with first experimental
setup after cooling in positive field). }\label{amplituden}
\end{figure}

With the first experimental setup on IN12 we investigated the
influence of the amplitude of the switching field by applying 1.2,
1.35, and 1.5\,kV/mm with a constant repetition rate of 8\ Hz, see
Fig.~\ref{amplituden}. The difference for the three applied
amplitudes can be easily analyzed: If we switch from the not
preferred state to the preferred state (i.e. the rapid process)
the field of 1.2\,kV/mm already largely exceeds the minimal
required field (i.\,e. the coercive field of the preferred state).
The maximal chirality and the stretching exponents stay almost the
same for all three fields, but the rise time diminishes with
increasing electric field. If we switch into the not-preferred
state more drastic changes occur as function of electric field.
The domain distribution reaches lower saturation values at smaller
fields, and, more importantly, the rise-times decrease more
rapidly than the inverse of the electric field, see Fig. 8. This
E-field dependence of the rise times allows us to overlay the
multiferroic domain behavior in the two distinct crystals.

\begin{figure}[t]
\includegraphics[width=0.75\columnwidth, angle=0]{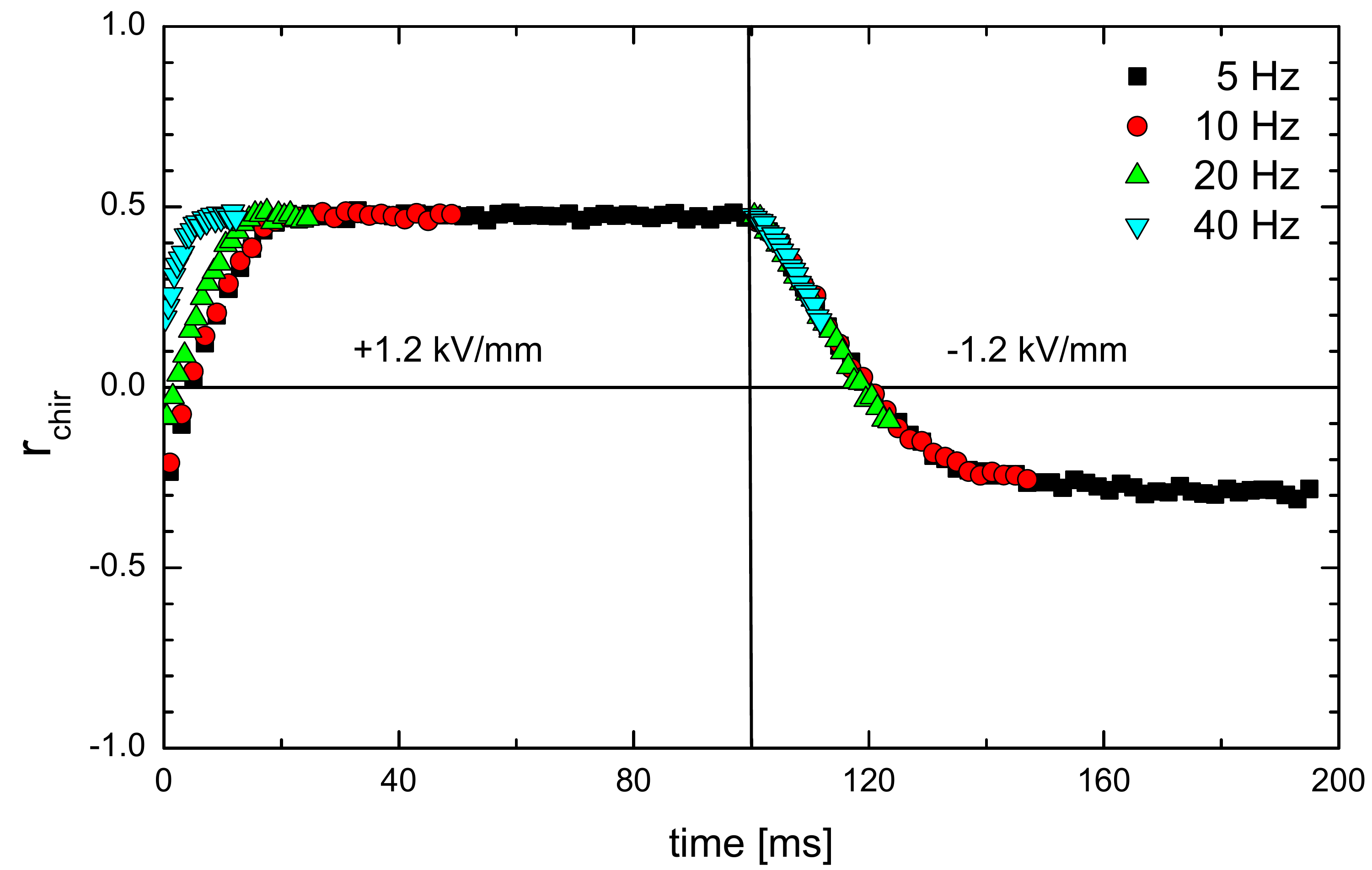}
\caption{Switching at different frequencies with the first setup.
At 10\,K the electric field is switched with different frequencies
and an amplitude of $\SI{\pm1.2}{kV/mm}$. The system clearly shows
a preferred state. At high frequencies the time is too short to
reach saturation in the not-preferred state. The data are shifted
in the time axis such that the time of the switch coincides for
the different frequencies (data taken with first experimental
setup after cooling in positive field).} \label{frequenz}
\end{figure}

In addition, we investigated the effect of the frequency on the
switching behavior. At 10\,K we switched at the following
frequencies (chronological order): 10, 12, 5, 8, 20, and 40\,Hz.
The measured curves at 5, 10, 20, and 40\,Hz are shown in
Fig.~\ref{frequenz}. The system behaves as expected for a
preferred state which is imprinted by the first cooling in strong
applied electric fields. When we switch faster than the system is
able to respond to, the saturation value of the chiral component
is not reached in that field direction. The dielectric studies
also observe incomplete domain switching depending on frequency
and temperature\cite{niermann}. The rise times in our experiment,
however, do not essentially change within the studied frequency
range.

During the entire sequence of experiments with the first setup we
realized that the sample exhibits some fatigue. We compare runs
which were done at 10\,K and 8\,Hz with different histories of the
sample. As reference we use the run which was taken in the
beginning of the measurement during the first temperature scan of
rise-time experiments. At this first time the chirality could be
flipped almost as large as in the static cycle (+0.49, $-0.45$) at
$T=\SI{10}{K}$, see Fig.~\ref{schalten}. Later, after the sample
had been switched with 12\,Hz for long time and after thermal
cycling, the preferred chirality almost stayed the same (+0.48)
but the not preferred chirality decreased to $-0.28$, see
Fig.~\ref{frequenz} for $\nu=\SI{5}{Hz}$. After taking the entire
frequency dependence, ending with switching at 40\ Hz, all
parameters stay more or less the same. This indicates a general
fatigue mechanism due to the switching and the thermal cycling
rather than  a destruction of the sample by applying a too high
frequency. After the observation of fatigue effects, we heated the
sample to 20\,K and cooled it again. However, the sample did not
restore to its initial behavior. So we heated the sample to 120\,K
in the hope that such procedure would restore the sample, however,
this was not accomplished. The repeated switching seems to induce
defects influencing magnetic and polar order which do not heal
when going far beyond the magnetic phase transition. With the
second setup no fatigue was observed, as described above.

\section{Conclusions}

In conclusion the time scales of controlling the magnetic chiral
components by an external electric field have been studied by
time-resolved polarized neutron diffraction experiments. By
comparing comprehensive measurements with two different samples
and with different poling history we can identify the general
features of the domain switching and those which depend on the
sample and on its thermal history.

In general the rise times to invert the magnetic components are
slow, typically of the order of a few to several tens of
milliseconds. This agrees with a measurement of second harmonic
generation and in particular with a comprehensive study by
dielectric spectroscopy. The magnetic response can be described by
single relaxation process (only in one case we find evidence for a
superposed even slower domain relaxation) and there is no
indication for a fast process. We always find a weakly enhanced
stretching exponent. The reversal of domains seem to be
essentially controlled by a single process type which preserves
the tight coupling between magnetic chirality and ferroelectric
polarization. The chiral magnetism cannot be reversed faster than
the ferroelectric polarization.

The peculiar temperature dependence of the rise times with a
maximum around 10\ K also is a general feature of the domain
kinetics in \mwo . The faster response is naturally expected when
approaching the upper phase transition, where ferroelectricity and
the chiral components disappear continuously associated with
softer excitations. However, the faster kinetics near the lower
first-order phase transition is astonishing and possibly related
to the increase of the anharmonic modulations upon cooling in the
multiferroic phase. The second- and higher-order modulations of
both, the nuclear and the crystal structure can be relevant for
the effective pinning of domains in \mwo . The independent
measurement on crystals of two different batches yield good
agreement concerning the qualitative temperature dependence and
even quantitative agreement when considering the average of the
four possibilities of the electric field (relaxation field and
initial poling) and taking the field strength into account.

The sign and the size of the asymmetries, which are seen in the
rise times, in the saturation chiral component or in the coercive
fields, clearly are sample and history dependent. Even after
zero-field cooling, a finite asymmetry persists indicating an
intrinsic property of the sample. Field cooling from higher
temperatures seems to imprint stronger polar preferences, which
can be reverted \cite{8} by different thermal cycling.

This work was supported by the Deutsche Forschungsgemeinschaft
through the Sonderforschungsbereich 608 and by the BMBF through
project  05K13PK1. We thank J.~Stein and J.~Hemberger for valuable
discussions.

\end{document}